\RequirePackage{fix-cm}
\documentclass[twocolumn,epjc3,final]{svjour3}  
\smartqed  
 \usepackage{mathrsfs}
\usepackage{amsmath, txfonts}
\usepackage{graphicx,bm,booktabs,bm}
\usepackage{longtable,lscape}
\usepackage{overpic,multirow,color}
\definecolor{cover}{rgb}{0.77,0.87,0.88}
\definecolor{blueone}{rgb}{0.1,0.1,.7}
\definecolor{citec}{rgb}{0.14,0.47,0.09}
\definecolor{two}{rgb}{0.0,0.5,0.}
\definecolor{three}{rgb}{.5,.1,0.15}
\usepackage[bookmarks=true,bookmarksopen=false,plainpages=false,breaklinks=true,
   bookmarksnumbered=true,hypertexnames=false,
   filecolor=blue,urlcolor=three,menucolor=three,
   linkcolor=three,citecolor=blueone, colorlinks,
   anchorcolor=blue,runcolor=pink,frenchlinks=red
   pdfstartview=FitH,pdftitle=title,%
   pdfauthor=author]{hyperref}
              \allowdisplaybreaks[4]

\graphicspath{{Figures/}} %

\journalname{Eur. Phys. J. C}
\begin{document}
\title{Study of $P_c(4457)$, $P_c(4440)$, and $P_c(4312)$ in a quasipotential Bethe-Salpeter equation approach}
\author{Jun He\thanksref{e1,addr1}
}                     
\thankstext{e1}{junhe@njnu.edu.cn}
\institute{Department of  Physics and Institute of Theoretical Physics, Nanjing Normal University,
Nanjing 210097, China \label{addr1}
}

\date{Received: date / Revised version: date}
%
\maketitle

\abstract{
Very recently, the LHCb Collaboration reported their new results about the pentaquarks at charm energy region. Based on the new  experimental results, we recalculate the molecular states composed of a $\Sigma_c^{(*)}$ baryon and a $\bar{D}^{(*)}$ meson in a quasipotential Bethe-Salpeter equation approach. The two-peak structure around 4450 MeV can be interpreted as two $\Sigma_c \bar{D}^*$ bound states with spin parities $1/2^-$ and $3/2^-$. The newly observed pentaquark $P_c(4312)$ can be assigned as a $\Sigma_c\bar{D}$ bound state with spin parity $1/2^-$. The experimental determination of  spin parities  will be very helpful to understand the internal structure of these pentaquarks.
} 

\section{Introduction}

In 2015, the LHCb reported their first observation of  pentaquark candidates, the $P_c(4450)$ and $P_c(4380)$, which carry puzzling opposite parities and large spins, especially best suggested value $J=5/2$ for the $P_c(4450)$~\cite{Aaij:2015tga}. The masses of $P_c(4450)$ and $P_c(4380)$ are very close to the $\Sigma_c \bar{D}^*$ and $\Sigma_c^* \bar{D}$ thresholds, respectively. It is very alluring to assign those two pentaquarks as corresponding molecular states. However, the puzzling opposite parties of these two pentaquarks force us to assign one of the state as a P-wave molecular state if we accept the suggested spin parities  of the experimentists. Very recently, the LHCb Collaboration reported their new results about the pentaquarks at the same energy region~\cite{LHCbtalk,Aaij:2019vzc}. It is very interesting to observe that the $P_c(4450)$ split into two peaks, $P_c(4457)$ and $P_c(4440)$, based on more accumulated data, and a new  pentaquark near the $\Sigma_c\bar{D}$ threshold was observed and named as $P_c(4312)$.

The new result is very helpful to deepen our understanding of the hidden-charm pentaquarks. Two higher states $P_c(4457)$ 
and $P_c(4440)$ correspond to original $P_c(4450)$ obviously. Though the partial wave analysis is on progress and their spin parities were not provided in the report, it is not surprising that the spin parities of the $P_c(4457)$ and $P_c(4440)$ will be different from the suggested spin parity $5/2^+$ of the $P_c(4450)$ in  previous experimental article~\cite{Aaij:2015tga}. Moreover, the two-peak structure  suggests that the large spin $5/2$ for the $P_c(4450)$ may be due to taking two peaks as one. Hence, these two peaks may be related to two S-wave $\Sigma_c \bar{D}^*$ molecular states with smaller spin, such as $1/2$ and $3/2$, and with negative parity. It is interesting to see that theoretically there exist only two S-wave  $\Sigma_c \bar{D}^*$ states and one S-wave $\Sigma_c\bar{D}$ state, which is consistent with the experimental observation of the $P_c(4457)$ and $P_c(4440)$ near  $\Sigma_c \bar{D}^*$ threshold and  $P_c(4312)$ near $\Sigma_c\bar{D}$ threshold, respectviely. 

In fact, in early predictions about the hidden-charm pentaquarks in the molecular state picture, the $\Sigma_c \bar{D}^*$ bound states were predicted with spin parities $1/2^-$ and $3/2^-$ ~\cite{Wang:2011rga,Yang:2011wz,Wu:2012md}. For example, in our early work~\cite{Yang:2011wz}, two isovector   $\Sigma_c \bar{D}^*$ bound state were predicted with $1/2^-$ and $3/2^-$ in the one-boson exchange model by solving the Sch\"odinger equation. However, the best suggested spin parity of the $P_c(4450)$ as $5/2^+$ gave a blow to such assignment. To interpret such spin parity, in our previous work~\cite{He:2015cea}, we assign the S-wave $3/2^-$ $\Sigma_c\bar{D}^*$  bound state  to the $P_c(4380)$ and the $P_c(4450)$ as a P-wave  $\Sigma_c\bar{D}^*$ bound state~\cite{He:2016pfa}. Even without attemp to give the opposite parities, the $P_c(4450)$ was assigned as a deeply-bound $\Sigma^*_c\bar{D}^*$ state and the $3/2^-$ $\Sigma_c\bar{D}^*$ state was left to interpret the $P_c(4380)$ in Ref.~\cite{Chen:2015loa}.  
In their new work~\cite{Chen:2019asm}, the $\Sigma^{(*)}_c\bar{D}^{(*)}$ system were restudied in the one-boson-exchange model by solving the Sch\"odinger equation. Three pentaquarks can be plausibly  reproduced as the molecular states near their corresponding thresholds with reasonable cutoffs. In Ref.~\cite{Chen:2019bip}, the possible assignment of these three states in the molecular state picture were also discussed  in the QCD sum rule approach.

Based on the new results, as discussed above, we expect that the $P_c(4457)$ and $P_c(4440)$ have smaller spins, such as $1/2$ and $3/2$. Moreover, the observation of  the $P_c(4312)$ near the $\Sigma_c\bar{D}$ threshold in the new experiment confirms the idea that these pentaquarks should be the molecular states if we notice that all three states are very narrow, which is a character of the molecular state.  In this work, we will study the $\Sigma^{(*)}_c\bar{D}^{(*)}$ system in a coupled-channel quasipotential Bethe-Salpeter approach to find the molecular states. We expect that with smaller  cutoff the $\Sigma_c\bar{D}^*$ abound state with $3/2^-$ in our previous works~\cite{He:2015cea,He:2016pfa} will move to its threshold and the  states with spin $5/2$ will vanish. And the existence of $\Sigma_c\bar{D}$ molecular state relevant to $P_c(4312)$ will be also studied in our approach.

This work is organized as follows. After introduction, the detail of the dynamical study of coupled-channel $\Sigma^{(*)}_c\bar{D}^{(*)}$  interactions will be presented, which includes relevant effective Lagrangian. The numerical results  are given in Section 3. Finally, summary  and  discussion will be given.

\section{Coupled-channel $\Sigma^{(*)}_c\bar{D}^{(*)}$  interactions and relevant Lagrangians}

In this work, we will consider coupled-channel interactions between a charmed baryon and an anti-charmed meson as $\Sigma_c\bar{D}-\Sigma_c^*\bar{D}-\Sigma_c\bar{D}^*-\Sigma^*_c\bar{D}^*$, which is described by the light meson exchanges.
We need to construct effective Lagrangians depicting the couplings of light mesons and  anti-charmed mesons or charmed baryons.

In terms of heavy quark limit and chiral symmetry, the couplings of light mesons (peseudoscalar $\mathbb{P}$, vector $\mathbb{V}$ and $\sigma$ mesons) to heavy-light charmed mesons $\tilde{\mathcal{P}}=(\bar{D}^0, D^-, D^-_s)$  are constructed  as \cite{Cheng:1992xi,Yan:1992gz,Wise:1992hn,Casalbuoni:1996pg},
\begin{eqnarray}\label{eq:lag-p-exch}
  \mathcal{L}_{\tilde{\mathcal{P}}^*\tilde{\mathcal{P}}\mathbb{P}} &=&
 i\frac{2g\sqrt{m_{\tilde{\mathcal{P}}} m_{\tilde{\mathcal{P}}^*}}}{f_\pi}
  (-\tilde{\mathcal{P}}^{*\dag}_{a\lambda}\tilde{\mathcal{P}}_b
  +\tilde{\mathcal{P}}^\dag_{a}\tilde{\mathcal{P}}^*_{b\lambda})
  \partial^\lambda\mathbb{P}_{ab},\nonumber\\
    \mathcal{L}_{\tilde{\mathcal{P}}^*\tilde{\mathcal{P}}^*\mathbb{P}} &=&
-\frac{g}{f_\pi} \epsilon_{\alpha\mu\nu\lambda}\tilde{\mathcal{P}}^{*\mu\dag}_a
\overleftrightarrow{\partial}^\alpha \tilde{\mathcal{P}}^{*\lambda}_{b}\partial^\nu\mathbb{P}_{ba},\nonumber\\
    \mathcal{L}_{\tilde{\mathcal{P}}^*\tilde{\mathcal{P}}\mathbb{V}} &=&
\sqrt{2}\lambda g_V\varepsilon_{\lambda\alpha\beta\mu}
  (-\tilde{\mathcal{P}}^{*\mu\dag}_a\overleftrightarrow{\partial}^\lambda
  \tilde{\mathcal{P}}_b  +\tilde{\mathcal{P}}^\dag_a\overleftrightarrow{\partial}^\lambda
  \tilde{\mathcal{P}}_b^{*\mu})(\partial^\alpha{}\mathbb{V}^\beta)_{ab},\nonumber\\
	\mathcal{L}_{\tilde{\mathcal{P}}\tilde{\mathcal{P}}\mathbb{V}} &=& -i\frac{\beta	g_V}{\sqrt{2}}\tilde{\mathcal{P}}_a^\dag
	\overleftrightarrow{\partial}_\mu \tilde{\mathcal{P}}_b\mathbb{V}^\mu_{ab}, \nonumber\\
  \mathcal{L}_{\tilde{\mathcal{P}}^*\tilde{\mathcal{P}}^*\mathbb{V}} &=& - i\frac{\beta
  g_V}{\sqrt{2}}\tilde{\mathcal{P}}_a^{*\dag}\overleftrightarrow{\partial}_\mu
  \tilde{\mathcal{P}}^*_b\mathbb{V}^\mu_{ab}\nonumber\\
  &-&i2\sqrt{2}\lambda  g_Vm_{\tilde{\mathcal{P}}^*}\tilde{\mathcal{P}}^{*\mu\dag}_a\tilde{\mathcal{P}}^{*\nu}_b(\partial_\mu\mathbb{V}_\nu-\partial_\nu\mathbb{V}_\mu)_{ab}
,\nonumber\\
  \mathcal{L}_{\tilde{\mathcal{P}}\tilde{\mathcal{P}}\sigma} &=&
  -2g_s m_{\tilde{\mathcal{P}}}\tilde{\mathcal{P}}_a^\dag \tilde{\mathcal{P}}_a\sigma, \nonumber\\
  \mathcal{L}_{\tilde{\mathcal{P}}^*\tilde{\mathcal{P}}^*\sigma} &=&
  2g_s m_{\tilde{\mathcal{P}^*}}\tilde{\mathcal{P}}_a^{*\dag}
  \tilde{\mathcal{P}}^*_a\sigma,
\end{eqnarray}
where constant $f_\pi=132$ MeV and $\mathbb
P$ and $\mathbb V$ are the pseudoscalar and vector matrices
\begin{eqnarray}
    {\mathbb P}&=&\left(\begin{array}{ccc}
        \frac{1}{\sqrt{2}}\pi^0+\frac{\eta}{\sqrt{6}}&\pi^+&K^+\\
        \pi^-&-\frac{1}{\sqrt{2}}\pi^0+\frac{\eta}{\sqrt{6}}&K^0\\
        K^-&\bar{K}^0&-\frac{2\eta}{\sqrt{6}}
\end{array}\right),\nonumber\\
\mathbb{V}&=&\left(\begin{array}{ccc}
\frac{\rho^{0}}{\sqrt{2}}+\frac{\omega}{\sqrt{2}}&\rho^{+}&K^{*+}\\
\rho^{-}&-\frac{\rho^{0}}{\sqrt{2}}+\frac{\omega}{\sqrt{2}}&K^{*0}\\
K^{*-}&\bar{K}^{*0}&\phi
\end{array}\right).
\end{eqnarray}

The effective Lagrangians depicting the charmed baryons with the light mesons with chiral symmetry, heavy quark limit and hidden local symmetry read \cite{Yan:1992gz,Liu:2011xc},
\begin{eqnarray}
{\cal L}_{\mathcal{B}\mathcal{B}\mathbb{P}}
&=&-\frac{g_1}{4f_\pi}~\epsilon^{\alpha\beta\lambda\kappa}
    \langle\bar{\mathcal{B}}~\overleftrightarrow{\partial}_\kappa\gamma_\alpha \gamma_\lambda
    \partial_\beta\mathbb{P} ~\mathcal{B}\rangle,\nonumber\\
{\cal L}_{\mathcal{B}\mathcal{B}\mathbb{V}} &=& -i\frac{\beta_S
g_{V}}{2\sqrt{2}}~\langle\bar{\mathcal{B}}~ \overleftrightarrow{\partial}\cdot
\mathbb{V}~ \mathcal{B}\rangle\nonumber\\
&&-\frac{im_{\mathcal{B}}\lambda_Sg_{V}}{3\sqrt{2}}~\langle\bar{\mathcal{B}}\gamma_\mu
\gamma_\nu
(\partial^\mu \mathbb{V}^\nu-\partial^\nu\mathbb{V}^\mu)\mathcal{B}\rangle,\nonumber\\
{\cal L}_{\mathcal{B}\mathcal{B}\sigma} &=&
-\ell_Sm_{\mathcal{B}}\langle\bar{\mathcal{B}}~\sigma~
\mathcal{B}\rangle,\label{ha1}\nonumber\\
{\cal L}_{\mathcal{B}^*\mathcal{B}^*\mathbb{P}}
&=&\frac{-3g_1}{4 f_\pi}~\epsilon^{\alpha\beta\lambda\kappa}
    \langle\bar{\mathcal{B}}^*_{\alpha}
   \overleftrightarrow{\partial}_\kappa \partial_\beta\mathbb{P} ~\mathcal{B}^*_{\lambda}\rangle,\nonumber\\
{\cal L}_{\mathcal{B}^*\mathcal{B}^*\mathbb{V}} &=& i\frac{\beta_S
	g_{V}}{2\sqrt{2}}~\langle\bar{\mathcal{B}}^{*\mu}\overleftrightarrow{\partial}\cdot
	\mathbb{V}\mathcal{B}^*_{\mu}\rangle\nonumber\\
	&&+\frac{im_{\mathcal{B}^*}\lambda_Sg_{V}}{\sqrt{2}}~\langle\bar{\mathcal{B}}^{*}_{\mu}
(\partial^\mu
\mathbb{V}^\nu-\partial^\nu\mathbb{V}^\mu)\mathcal{B}^*_{\nu}\rangle,\nonumber\\
{\cal L}_{\mathcal{B}^*\mathcal{B}^*\sigma} &=&
\ell_Sm_{\mathcal{B}^*}\langle\bar{\mathcal{B}}^*~\sigma~
\mathcal{B}^*\rangle,\nonumber\\\label{ha1}
{\cal L}_{\mathcal{B}\mathcal{B}^*\mathbb{P}}
&=&\frac{\sqrt{3}g_S}{4f_\pi}~\epsilon^{\mu\nu\lambda\kappa} ~\langle\bar{\mathcal{B}}\partial_\nu\mathbb{P}\overleftrightarrow{\partial}_\kappa\gamma^5\gamma_\mu
\mathcal{B}^*_\lambda-\bar{\mathcal{B}}^*_\mu\partial_\nu\mathbb{P}\overleftrightarrow{\partial}_\kappa\gamma_\lambda\gamma^5
\mathcal{B}\rangle,\nonumber\\
{\cal L}_{\mathcal{B}\mathcal{B}^*\mathbb{V}}&=&\frac{\beta_S g_V}{4\sqrt{6m_{\mathcal{B}}m_{\mathcal{B}^*}}}
\langle\bar{\mathcal{B}}\overleftrightarrow{\partial}\cdot\mathbb{V}\overleftarrow{\partial}^\mu
\gamma^5\mathcal{B}^*_\mu
+\bar{\mathcal{B}}^*_\mu\gamma^5\overleftrightarrow{\partial}\cdot\mathbb{V}
\overrightarrow{\partial}^\mu \mathcal{B}\rangle\nonumber\\
&+&\frac{i\lambda_S
g_V\sqrt{m_{\mathcal{B}}m_{\mathcal{B}^*}}}{\sqrt{6}}\langle\bar{\mathcal{B}}\gamma^5(\partial^\mu \mathbb{V}^\nu-\partial^\nu \mathbb{V}^\mu)(\gamma_\nu+\frac{i\overleftrightarrow{\partial}_\nu}{2\sqrt{m_{\mathcal{B}}m_{\mathcal{B}^*}}})\mathcal{B}^*_\mu\nonumber\\
&-&\bar{\mathcal{B}}^*_\mu
(\partial^\mu \mathbb{V}^\nu-\partial^\nu \mathbb{V}^\mu)(\gamma_\nu+
\frac{i\overleftrightarrow{\partial}_\nu}{2\sqrt{m_{\mathcal{B}}m_{\mathcal{B}^*}}})\gamma^5\mathcal{B}\rangle
\label{LB}
\end{eqnarray}
where the partial $\overleftrightarrow{\partial}$ operates on the initial and final baryons and  $\mathcal{B}$ matrix is
\begin{eqnarray}
\mathcal{B}&=&\left(\begin{array}{ccc}
\Sigma_c^{++}&\frac{1}{\sqrt{2}}\Sigma^+_c&\frac{1}{\sqrt{2}}\Xi'^+_c\\
\frac{1}{\sqrt{2}}\Sigma^+_c&\Sigma_c^0&\frac{1}{\sqrt{2}}\Xi'^0_c\\
\frac{1}{\sqrt{2}}\Xi'^+_c&\frac{1}{\sqrt{2}}\Xi'^0_c&\Omega^0_c
\end{array}\right).
\end{eqnarray}
The $\langle \cdots\rangle$ in Eq.~(\ref{LB}) denotes the trace over the the $3\times3$ matrices.

To constrain the Lagrangians, the coupling constants should be determined. The values used in the calculation are listed in Table.
\ref{coupling}, which are from the literature~\cite{Chen:2019asm,Liu:2011xc,Isola:2003fh,Falk:1992cx}.
\renewcommand{\arraystretch}{1.5}
\begin{table}[h!]
\caption{The parameters and coupling constants adopted in our
calculation. The $\lambda$ and $\lambda_S$ are in the unit of GeV$^{-1}$. Others are in the unit of $1$.
\label{coupling}}
\begin{tabular}{cccccccccccccccccc}\hline
$\beta$&$g$&$g_V$&$\lambda$ &$g_{_S}$&$\beta_S$&$\ell_S$&$g_1$&$\lambda_S$\\
0.9&0.59&5.8&0.56 &0.76&-1.74&6.2&0.94&-3.31\\
\hline
\end{tabular}
\end{table}

With  above Lagrangians, the vertices for the heavy meson/baryon and the exchanged light meson can be obtained. The potential for the coupled channel interaction can be written with these vertices and the propagator of the exchanged light meson with the help of the Feynman rules.  Here the propagators for the exchanged mesons read,
\begin{eqnarray}%
P_\mathbb{P}(q^2)&=&i\left(-\frac{1}{q^2-m_\pi^2}+\frac{1}{6}\frac{1}{q^2-m_\eta^2}\right),\nonumber\\
P^{\mu\nu}_\mathbb{V}(q^2)&=&i\left(-\frac{(-g^{\mu\nu}+q^\mu q^\nu/m^2_{\rho})}{q^2-m_\rho^2}-\frac{1}{2}\frac{(-g^{\mu\nu}+q^\mu q^\nu/m^2_{\omega})}{q^2-m_\omega^2}\right),\nonumber\\
P_\sigma(q^2)&=&\frac{i}{q^2-m^2_\sigma}.
\end{eqnarray}
Because four channels are considered in the current work, it is 
tedious and fallible  to give the explicit of  16 potential elements for the coupled-channel interaction and input them into the code.  Instead, in this work, the vetices $\Gamma$ and the above propagators $P$ will be input in to the code directly and the potential can be obtained as
\begin{eqnarray}%
{\cal V}_{\mathbb{P},\sigma}=\Gamma_1\Gamma_2 P_{\mathbb{P},\sigma}(q^2),\quad
{\cal V}_{\mathbb{V}}=\Gamma_{1\mu}\Gamma_{2\nu}  P^{\mu\nu}_{\mathbb{V}}(q^2).
\end{eqnarray}
Hence, we do not give the explicit forms of the potentials here.

As our previous work~\cite{He:2015cea}, the Bethe-Saltpeter equation approach with a spectator quasipotential approximation, which was explained explicitly in the appendices of Ref.~\cite{He:2015mja}, will be adopted to search the possible bound states. A bound state from the interaction corresponds to a pole of the scattering amplitude ${\cal M}$  which is described by potential kernel obtained in the above. The Bethe-Saltpeter equation for partial-wave amplitude with fixed spin-parity $J^P$ reads ~\cite{He:2015mja},
\begin{eqnarray}
i{\cal M}^{J^P}_{\lambda'\lambda}({\rm p}',{\rm p})
&=&i{\cal V}^{J^P}_{\lambda',\lambda}({\rm p}',{\rm
p})+\sum_{\lambda''}\int\frac{{\rm
p}''^2d{\rm p}''}{(2\pi)^3}\nonumber\\
&\cdot&
i{\cal V}^{J^P}_{\lambda'\lambda''}({\rm p}',{\rm p}'')
G_0({\rm p}'')i{\cal M}^{J^P}_{\lambda''\lambda}({\rm p}'',{\rm
p}).\quad\quad \label{Eq: BS_PWA}
\end{eqnarray}
Note that the sum extends only over nonnegative helicity $\lambda''$.
The partial wave potential is defined as
\begin{eqnarray}
{\cal V}_{\lambda'\lambda}^{J^P}({\rm p}',{\rm p})
&=&2\pi\int d\cos\theta
~[d^{J}_{\lambda\lambda'}(\theta)
{\cal V}_{\lambda'\lambda}({\bm p}',{\bm p})\nonumber\\
&+&\eta d^{J}_{-\lambda\lambda'}(\theta)
{\cal V}_{\lambda'-\lambda}({\bm p}',{\bm p})],
\end{eqnarray}
where $\eta=PP_1P_2(-1)^{J-J_1-J_2}$ with $P$ and $J$ being parity and spin for system, $\bar{D}^{(*)}$ meson or $\Sigma_c^{(*)}$ baryon. The initial and final relative momenta are chosen as ${\bm p}=(0,0,{\rm p})$  and ${\bm p}'=({\rm p}'\sin\theta,0,{\rm p}'\cos\theta)$ with ${\rm p}^{(')}=|{\bm p}^{(')}|$. The $d^J_{\lambda\lambda'}(\theta)$ is the Wigner d-matrix. 
In the chiral unitary approach, an cutoff in momentum  was introduced and can be seen
as regularization~\cite{Oset:1997it}, which is
related to the dimensional regularization as discussed in Ref.~\cite{Oller:1998hw}.  In this work we will adopt an  exponential
regularization by introducing a form factor in the propagator as
\begin{eqnarray}
G_0({\rm p})\to G_0({\rm p})\left[e^{-(k_1^2-m_1^2)^2/\Lambda^4}\right]^2,\label{regularization}
\end{eqnarray} 
with $k_1$ and $m_1$ being the momentum and mass of the charmed meson.  The interested reader is referred to Ref.~\cite{He:2015mja} for further information about the regularization. The value of the cutoff $\Lambda$ should be about 1 GeV. 
 A monopole form factor is introduced to compensate the
off-shell effect of exchanged meson  as  $f(q^2)=({\Lambda_e^2-m^2})/({\Lambda_e^2-q^2})$ with cutoff $\Lambda_e=m+\alpha~0.22$ GeV. Here $m$ and $q$ are the mass and momentum of the exchanged light meson. In this work, we fix the $\alpha=2$. 

After transforming the integral equation to a matrix equation, the pole of scattering  amplitude $\cal M$ can be searched by variation of $z$ to satisfy
$|1-V(z)G(z)|=0$
with  $z=M+i\Gamma/2$ equaling to the system energy $M$ at the
real axis~\cite{He:2015mja}.

\section{Numerical results}

In this work, the only free parameter is the cutoff $\Lambda$ in the exponential regularization in Eq.~(\ref{regularization}). We will vary the cutoff $\Lambda$ from 0.8
to 1.8 GeV to search for the poles which correspond to  bound states
from the interactions in an energy region 4.3 GeV to 4.5 GeV where the $P_c(4457)$, $P_c(4440)$ and $P_c(4312)$ were observed. The coupled-channel $\Sigma_c\bar{D}-\Sigma_c^*\bar{D}-\Sigma_c\bar{D}^*-\Sigma^*_c\bar{D}^*$ interaction is considered in the current calculation.  Three poles can be found and listed in Tables~\ref{1/2} and~\ref{3/2}.
\renewcommand\tabcolsep{0.267cm}
\renewcommand{\arraystretch}{1.5}
\begin{table}[h!]
\caption{The position of the bound states with $J^P=1/2^-$ in a unit of MeV. The $\Lambda$ in the unit of GeV is the cutoff in the exponential regularization in Eq.~(\ref{regularization}). The $CC$ means coupled-channel calculation and $\Sigma_c\bar{D}^*$ and $\Sigma_c\bar{D}$ mean the corresponding single-channel calculation.
\label{1/2}}
\begin{tabular}{cccccccccccccccccc}\hline
$\Lambda$&$CC(1/2^-)_1$&$\Sigma_c\bar{D}^*(1/2^-)$ &$CC(1/2^-)_2$&$\Sigma_c\bar{D}(1/2^-)$  \\
\hline
0.8 &4448.0 &4450.8 &4320.4  &4320.3 \\
1.0 &4439.4 &4444.8 &4320.0  &4319.5\\
1.2 &4431.7 &4439.0 &4320.1  &4318.7\\
1.4 &4425.2 &4433.9 &4320.2  &4318.0 \\
1.6 &4420.1 &4429.8 &4320.3  &4317.5\\
1.8 &4416.4 &4426.7 &4320.5  &4317.1\\
\hline
\end{tabular}
\end{table}

Form Table~\ref{1/2}, one can find that two poles can be produced from the coupled-channel interaction with spin parity $J^P=1/2^-$. The upper state has a mass of about 4440 MeV, and the mass will decrease with the increase of the cutoff. The lower state has a mass of about 4320 MeV, which is considerably stable with the variation of the cutoff. For comparison, we also provide the results with single-channel calculation for the $\Sigma_c\bar{D}^*$ and the $\Sigma_c\bar{D}$ interactions. Generally speaking, the coupled-channel  effect is considerable small, which is consistent with the results for coupled-channel $\Sigma^*K-\Sigma K^*$ interaction~\cite{He:2015yva}. The results suggest that the upper and lower states from the coupled-channel calculation can be taken as the $\Sigma_c\bar{D}^*$ and $\Sigma_c\bar{D}$ molecular states with $1/2^-$.

\renewcommand\tabcolsep{0.9cm}
\renewcommand{\arraystretch}{1.5}
\begin{table}[h!]
\caption{The position of the bound states with $J^P=3/2^-$ in a unit of MeV. The $\Lambda$ in the unit of GeV is the cutoff in the exponential regularization. The $CC$ means coupled-channel calculation and $\Sigma_c\bar{D}^*$ means the single-channel calculation of $\Sigma_c\bar{D}^*$ interaction.
\label{3/2}}
\begin{tabular}{cccccccccccccccccc}\hline
$\Lambda$ &$CC(3/2^-)$&$\Sigma_c\bar{D}^*(3/2^-)$  \\
\hline
0.8 & $--$  & 4462.0 \\
1.0 &4461.7 &4461.9\\
1.2 &4461.5 &4461.7\\
1.4 &4461.3 &4461.6\\
1.6 &4461.1 &4461.4\\
1.8 &4460.9 &4461.3\\

\hline
\end{tabular}
\end{table}

In the case of spin parity $3/2^-$, only one state is found at about 4460 MeV as shown in Table~\ref{3/2}. We also compare the coupled-channel results with the single-channel $\Sigma_c\bar{D}^*$ calculation. The results suggest this state with $3/2^-$ is mainly from the $\Sigma_c\bar{D}^*$ interaction. As the two $1/2^-$ states, this state is also not sensitive  on the cutoff. Though there is no bound state from the $\Sigma_c^*\bar{D}$ channel,  an enhancement can be observed near the $\Sigma_c^*\bar{D}$ threshold, which suggests the interaction in this channel is still attractive.

\section{Summary and discussion}

Inspired by new LHCb results about the hidden-charm pentaquarks. We reanalyze the $\Sigma^{(*)}_c\bar{D}^{(*)}$  interaction in the quasipotential Bethe-Salpeter equation approach. In our previous works~\cite{He:2015cea,He:2016pfa}, the $P_c(4450)$ is interpreted as a $5/2^+$  $\Sigma_c\bar{D}^*$ state. In the current work, we persist in the assumption that the hidden-charm pentaquarks are from the molecular state with the nearest threshold, that is, two new pentaquarks with masses of about 4450 MeV are still expected to be the $\Sigma_c\bar{D}^*$ molecular states as we did for the $P_c(4450)$ while we do not force them to have a large spin $5/2$. The current results suggest that two states with spin parities $1/2^-$ and  $3/2^-$ can be produced from the coupled-channel calculation and mainly from the $\Sigma_c\bar{D}^*$ interaction, which can be assigned as the $P_c(4440)$ and $P_c(4457)$, respectively. It is consistent with our original work in Ref.~\cite{Yang:2011wz} and recent calculation in Ref.~\cite{Chen:2019asm}. 

The newly observed pentaquark $P_c(4312)$ can be assigned as a $1/2^-$ $\Sigma_c\bar{D}$ molecular state, which is the lower state in our coupled-channel calculation. In the current work, no bound state from $\Sigma^*\bar{D}$ interaction can be found while the interaction is still attractive  with the cutoff adopted. The $P_c(4380)$ is a broad state in the old experimental analysis~\cite{Aaij:2015tga}. In the new experimental results of $J/\psi p$ invariant mass spectrum near the  $\Sigma^*\bar{D}$ threshold, there still does not exist obvious sharp peak~\cite{LHCbtalk}. However, small structure can not be excluded before the partial-wave analysis being made. The weaker but attractive $\Sigma^*\bar{D}$ interaction may lead to  smaller and broader structure around 4380 GeV.

The report by LHCb collaboration does not provide the information of the spin parities of these three pentaquarks. So, other interpretations can not be excluded directly. For example, the two upper states can still be explained as $5/2^+$ and $5/2^-$  $\Sigma_c\bar{D}^*$  molecular state as suggested in Refs.~\cite{He:2015cea,He:2016pfa}, and the $P_c(4312)$ is a deeply bound $\Sigma^*_c\bar{D}$ state. In Ref.~\cite{Chen:2019bip}, the assignments of these three pentaquarks are also different from the suggestion in the current work and Ref.~\cite{Chen:2015loa}.   Hence, the partial wave analysis in experiment is essential to understand the origin and the internal structure of these pentaquarks.

\vskip 10pt

\noindent {\bf Acknowledgement} This project is supported by the National Natural Science
Foundation of China (Grants No. 11675228).

%


\end{document}